%% file: main.tex
\documentclass{article}

\usepackage{arxiv}

\usepackage[utf8]{inputenc} 
\usepackage[T1]{fontenc}    

\usepackage{amsmath}
\usepackage{amssymb}
\usepackage{amsfonts}

\usepackage{graphicx}
\usepackage{float}
\usepackage{placeins}
\usepackage{needspace}

\usepackage{booktabs}
\usepackage{multirow}
\usepackage{array}

\usepackage{algorithm}
\usepackage{algorithmic}

\usepackage[numbers,sort&compress]{natbib}
\usepackage{url}
\usepackage{microtype}
\usepackage[hidelinks]{hyperref}
\usepackage{doi}
\usepackage{cleveref}
\usepackage{etoolbox}

\title{Quadratic Unconstrained Binary Optimization for Sparse Magnetoencephalography Source Localization}

\date{August 13, 2026}

\newcommand{\orcid}[1]{%
  \href{https://orcid.org/#1}{%
    \IfFileExists{orcid.pdf}
      {\includegraphics[scale=0.06]{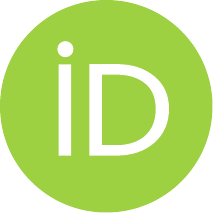}\hspace{1mm}}
      {\textsuperscript{\textsc{orcid}}\hspace{1mm}}%
  }%
}

\author{%
\orcid{0009-0008-8181-5055}Arim Ryou \\
Department of Physics, \\ Chungbuk National University \\
Cheongju 28644, Republic of Korea \\
\texttt{arimryou@chungbuk.ac.kr} \\
\And
\orcid{0000-0003-1195-5681}Kiwoong Kim%
\thanks{Corresponding author: \texttt{kiwoong@chungbuk.ac.kr}} \\
Department of Physics, \\ Chungbuk National University \\
Cheongju 28644, Republic of Korea \\
\texttt{kiwoong@chungbuk.ac.kr}
}
\renewcommand{\headeright}{}
\renewcommand{\undertitle}{}
\renewcommand{\shorttitle}{QUBO for Sparse Magnetoencephalography Source Localization}

\hypersetup{
pdftitle={Quadratic Unconstrained Binary Optimization for Sparse Magnetoencephalography Source Localization},
pdfsubject={Magnetoencephalography source localization using quadratic unconstrained binary optimization},
pdfauthor={Arim Ryou, Kiwoong Kim},
pdfkeywords={magnetoencephalography, source localization, quadratic unconstrained binary optimization, sparse support selection, inverse problems}
}

\graphicspath{{figures/}}
\begin{document}
\maketitle

\begin{abstract}
Magnetoencephalography (MEG) source localization is an ill-posed inverse
problem because distinct cortical source configurations can produce similar
sensor-level fields. We formulate sparse multi-source localization as a quadratic unconstrained
binary optimization (QUBO) problem combined with residual-aware candidate
screening. Candidate source-location groups are generated from the sensor-space
residual, fixed sensor-space templates are estimated for the resulting
candidates, and active templates are jointly selected using data-fit,
pairwise template interactions, and soft-cardinality terms. We evaluate the
method using classical simulated annealing in controlled synthetic MEG
simulations, primarily under a two-source condition, and compare it with
MNE, dSPM, MxNE, LCMV, and RAP-MUSIC. Across 100 main-benchmark trials,
QUBO achieved a mean cardinality-aware localization error of
\(8.45\,\mathrm{mm}\), compared with \(22.35\,\mathrm{mm}\) for MxNE, the
best-performing baseline according to this metric, corresponding to a
\(62.2\%\) reduction. The composite metric adds a \(50\,\mathrm{mm}\)
penalty per unit of source-count mismatch before normalization by the true
source count. Because MxNE returned only one source in 35 trials, the
reported reduction reflects both spatial localization and source-count
performance. In separate sensitivity experiments, QUBO remained competitive
across the tested sensor-noise and source-count conditions, although
RAP-MUSIC performed comparably to or better than QUBO in some low-noise and
three-source settings. The present experiments use classical simulated annealing
and do not evaluate quantum hardware or claim quantum advantage. The resulting
binary quadratic objective admits a direct Ising representation, enabling
future evaluation on quantum-annealing and hybrid backends.
\end{abstract}


\input{sections/01_introduction}
\input{sections/02_background}
\input{sections/03_method}
\input{sections/04_evaluation}
\input{sections/05_results}

\input{sections/06_discussion}
\input{sections/07_conclusion}

\bibliographystyle{unsrtnat}
\bibliography{references}


\end{document}

%% file: sections/01_introduction.tex
\section{Introduction}
\label{sec:introduction}

Magnetoencephalography (MEG) and electroencephalography (EEG), collectively
referred to as M/EEG, are noninvasive neuroimaging modalities that record,
respectively, magnetic fields and electric potentials generated by neural
activity with millisecond temporal resolution. A central problem in M/EEG
analysis is source localization, namely estimating the cortical locations that
generated the signals observed at the sensors. This problem is intrinsically
ill posed: the number of candidate cortical source locations is much larger
than the number of sensors, and distinct source configurations can produce
nearly indistinguishable sensor measurements. Stable source localization
therefore requires regularization or structural assumptions
\citep{Hamalainen1993,Baillet2001}.

Existing source-localization methods address this ill-posedness from different
perspectives. Minimum-norm methods estimate distributed source maps through
regularization of source amplitudes or source energy \citep{Hamalainen1994},
while dSPM and sLORETA apply statistical or localization-oriented normalization
to such estimates \citep{Dale2000,PascualMarqui2002}. LCMV beamforming
constructs spatial filters for candidate locations \citep{VanVeen1997}, and
subspace methods such as MUSIC and RAP-MUSIC exploit the low-dimensional
structure of sensor measurements \citep{Mosher1992,Mosher1999}. Sparse
representation methods, including \(\ell_1\)-regularization, mixed-norm
regularization, and sparse Bayesian learning, promote solutions supported on
relatively few cortical locations
\citep{Uutela1999,Ou2009,Wipf2009,Gramfort2012}. Although sparsity is not an
appropriate assumption for every form of neural activity, it provides a useful
modeling perspective in evoked-response and focal-activation settings in which
a limited number of cortical regions account for a substantial portion of the
measured sensor signal.

These methods also differ in how active source support is obtained. Distributed
and sparse-amplitude methods produce estimates in different forms and may
require an additional selection rule when a fixed number of representative
source locations is needed. Subspace methods such as RAP-MUSIC return discrete
dipole locations but construct the support sequentially by adding candidates
one at a time \citep{Mosher1999}. In multi-source settings, simultaneously
active sources are mixed at the sensors, and early selections can therefore
affect subsequent estimates. This motivates a formulation in which
candidate-specific contributions are evaluated and selected jointly under a
single support-selection objective.

Annealing-based global optimization has previously been investigated for MEG
and EEG dipole localization. For example, a hybrid approach combining simulated
annealing with a quasi-Newton local search has been evaluated for simultaneous
MEG--EEG dipole localization under low signal-to-noise-ratio conditions
\citep{Bastola2024}. The contribution of the present work therefore lies not in
the use of simulated annealing itself, but in the residual-aware construction
of candidate-specific sensor-space templates and their joint binary selection
through a soft-cardinality QUBO objective.

In this paper, we formulate sparse M/EEG source localization as a binary
source-support selection problem. Each candidate cortical source-location group
is assigned a binary variable indicating whether it belongs to the active
support, and support selection is expressed as a QUBO problem \citep{Kochenberger2014,Glover2018}. The proposed
objective is derived from a sensor-space residual-energy criterion induced by
the M/EEG forward model. Candidate-specific sensor-space contributions are
estimated through regularized linear fitting, and their individual data-fit
terms, pairwise template interactions, and cardinality control are combined in
a single binary quadratic objective. Target source-count information is
incorporated through a quadratic soft-cardinality penalty rather than imposed
as a hard equality constraint. The resulting model is therefore a
domain-specific QUBO formulation for M/EEG source-support selection.

A central motivation for adopting the QUBO representation is the combinatorial
growth of the source-support search space. For \(N\) candidate groups, the
binary search space contains \(2^N\) possible supports; even when the number of
active sources is fixed at \(k\), exact enumeration requires consideration of
\(\binom{N}{k}\) supports. Although a QUBO formulation does not by itself
guarantee a computational speedup, it provides a common representation
compatible with annealing-based optimization frameworks. Through the affine
transformation \(z_i=(1+s_i)/2\), where \(s_i\in\{-1,+1\}\), the QUBO objective
can be expressed as an equivalent Ising energy function \citep{Lucas2014}.
Such representations can be addressed using classical, quantum, and
quantum--classical hybrid annealing approaches. In the present study, we
evaluate the formulation using only classical simulated annealing. Its
performance on physical quantum-annealing hardware, computational scaling
benefits, and potential quantum advantage remain outside the scope of this
study.

Because constructing a QUBO over the full cortical source space would produce
a large binary optimization problem with a quadratic number of potential
pairwise terms, we use residual-aware candidate screening to construct a
reduced QUBO instance. Source groups are ranked according to the reduction in
residual energy obtained through ridge-regularized fitting, and the residual is
iteratively updated so that candidates explaining signal components not
represented by earlier selections can be included. Candidate-specific
sensor-space templates are constructed from successively updated residuals.
The final binary support is then selected jointly within this sequentially
constructed candidate and template set, after which continuous coefficients
are jointly refitted on the selected source groups.

We evaluate the proposed method using classical simulated annealing on
controlled synthetic MEG source-localization benchmarks for which ground-truth
source locations are available. Although the formulation is applicable to both
MEG and EEG forward models, the empirical evaluation in this study is limited
to synthetic MEG data. The primary objective is accurate identification of
active cortical locations rather than complete waveform reconstruction or
source-amplitude estimation. Target source-count information is assumed to be
available as prior information, with the primary benchmark conducted under a
two-source condition. Separate experiments examine the effects of sensor noise,
the number of active sources, candidate coverage, and alternative cardinality
formulations. Validation on real MEG data is outside the scope of this study
and is discussed as a limitation and future direction in
Section~\ref{sec:discussion}.

The contributions of this work are threefold. First, we derive a domain-specific
QUBO formulation for sparse M/EEG source-support selection from a sensor-space
residual-energy criterion, incorporating candidate-wise data-fit contributions,
pairwise template interactions, and soft cardinality control. Second, we
develop a sequential residual-aware candidate-screening and
template-construction procedure, followed by joint binary support selection
within the resulting reduced set. Third, we evaluate the formulation using
classical simulated annealing on controlled synthetic MEG benchmarks, compare
it with representative source-localization baselines, and analyze the effects
of sensor noise, source count, candidate coverage, and cardinality formulation.
These experiments characterize the feasibility and localization behavior of
the proposed formulation but do not assess computational scaling benefits or
quantum advantage.

The remainder of the paper is organized as follows.
Section~\ref{sec:background} introduces the M/EEG forward model, sparse support
selection, and the QUBO formulation. Section~\ref{sec:method} presents the
proposed method and its design rationale. Section~\ref{sec:evaluation}
describes the simulation benchmark, evaluation metrics, and baseline methods.
Section~\ref{sec:results} reports the experimental results.
Section~\ref{sec:discussion} discusses the implications and limitations of the
method, and Section~\ref{sec:conclusion} concludes the paper.

%% file: sections/02_background.tex
\section{Background and Problem Setup}
\label{sec:background}

This section introduces the M/EEG source-localization problem and the binary
support-selection perspective adopted in this work. We first describe the
linear forward model and the ill-posed nature of the corresponding inverse
problem. We then relate conventional sparse source estimation to binary
support selection at the cortical source-location level. Finally, we introduce
the general QUBO form used in the proposed method.

\subsection{M/EEG Forward Model and Inverse Problem}
\label{subsec:forward_inverse_problem}

M/EEG source localization aims to estimate neural current activity in the
cerebral cortex from MEG or EEG sensor measurements. The relationship between
sensor-level measurements and cortical source activity is commonly modeled
using the linear forward model
\citep{Hamalainen1993,Baillet2001,Gramfort2014}:
\begin{equation}
    Y_{\mathrm{raw}} = G_{\mathrm{raw}} J + E .
    \label{eq:raw_forward_model}
\end{equation}

Here, \(Y_{\mathrm{raw}} \in
\mathbb{R}^{n_{\mathrm{ch}} \times T}\) denotes the observed sensor data,
\(G_{\mathrm{raw}} \in
\mathbb{R}^{n_{\mathrm{ch}} \times n_{\mathrm{dof}}}\) is the lead-field or
gain matrix, \(J \in
\mathbb{R}^{n_{\mathrm{dof}} \times T}\) is the source-current matrix, and
\(E \in \mathbb{R}^{n_{\mathrm{ch}} \times T}\) denotes additive measurement
noise. The quantities \(n_{\mathrm{ch}}\), \(T\), and \(n_{\mathrm{dof}}\)
denote the numbers of sensors, time samples, and source degrees of freedom,
respectively. In a free-orientation model, \(n_{\mathrm{dof}}\) includes
multiple orientation components for each cortical source location.

The corresponding inverse problem is to estimate the source-current matrix
\(J\), or to identify the set of active cortical source locations, from the
observed data \(Y_{\mathrm{raw}}\) and the precomputed gain matrix
\(G_{\mathrm{raw}}\). This inverse problem is intrinsically ill posed: in
typical M/EEG settings, the number of source degrees of freedom is much larger
than the number of sensors, and distinct source configurations can produce
similar sensor-level measurements
\citep{Hamalainen1993,Baillet2001}. Consequently, the sensor data alone cannot
uniquely determine the underlying source activity, and stable estimation
requires regularization or additional structural assumptions.

In this work, the primary objective is to identify the active source support
rather than to use a distributed source map as the final localization output.
Prior information about a target number of active sources \(k\) is assumed to
be available. In the primary formulation, this information is incorporated
through a soft cardinality penalty rather than imposed as a hard equality
constraint.

In the subsequent formulation, \(Y\) denotes the sensor data obtained after
sensor-space whitening and localization time-window selection, and \(G\)
denotes the correspondingly whitened gain matrix after group-wise depth
normalization. For source-location group \(i\), \(G_i\) denotes the
corresponding block of the preprocessed gain matrix. These preprocessing and
source-grouping operations are described in Section~\ref{sec:method}.

\subsection{From Sparse Source Estimation to Binary Support Selection}
\label{subsec:sparse_to_binary_support}

A common approach to sparse M/EEG source localization is to estimate a
source-current matrix that explains the observed sensor data while promoting
a small or structured active support. Sparse M/EEG inverse methods include
minimum-current estimation, \(\ell_1\)-regularized formulations, mixed-norm
estimation, and sparse Bayesian learning
\citep{Uutela1999,Ou2009,Wipf2009,Gramfort2012}. Using the preprocessed sensor
data and gain matrix, a generic sparse source-estimation problem can be written
as
\begin{equation}
    \min_{J}
    \frac{1}{2}\|Y-GJ\|_F^2+\eta\mathcal{R}(J),
    \label{eq:sparse_source_estimation}
\end{equation}
where \(\mathcal{R}(J)\) promotes sparsity or structured sparsity and
\(\eta>0\) controls the trade-off between data fidelity and regularization.

The support considered in this work is defined at the cortical
source-location level. Let \(J_i \in \mathbb{R}^{d_i \times T}\) denote the
coefficient block associated with source-location group \(i\), where \(d_i\)
is the number of orientation components assigned to that location. In a
fixed-orientation model, \(d_i=1\), and \(J_i\) contains the source time course
associated with one gain column. In a free-orientation model, \(J_i\) contains
the orientation-component time courses associated with one cortical location.
For a block-sparse source estimate \(J\), the active source support is defined
as
\begin{equation}
    \mathcal{S}(J)
    =
    \left\{
        i : \|J_i\|_F > 0
    \right\}.
    \label{eq:active_source_support}
\end{equation}

Rather than treating the continuous estimate in
Equation~\eqref{eq:sparse_source_estimation} as the final localization output,
we focus on identifying the source-location groups that belong to the active
support. Let \(N\) denote the number of source-location groups included in a
given support-selection problem. We introduce the binary variables
\begin{equation}
    z_i =
    \begin{cases}
        1, & \text{if source-location group \(i\) is selected},\\
        0, & \text{otherwise},
    \end{cases}
    \qquad i=1,\ldots,N .
    \label{eq:binary_source_support}
\end{equation}

The target source count \(k\) is used as prior information to control the
cardinality of the selected support. It does not require the optimized support
to contain exactly \(k\) groups. In the proposed method, the binary variables
are defined over the reduced candidate set \(\mathcal{C}\) constructed before
QUBO optimization, such that \(N=|\mathcal{C}|\), rather than over the full
cortical source space. This binary representation leads naturally to the QUBO
formulation introduced in the next subsection.

\subsection{QUBO for Binary Support Selection}
\label{subsec:qubo_binary_support_selection}

QUBO is a standard binary quadratic optimization form used to represent combinatorial
problems involving both individual selection terms and pairwise couplings
\citep{Kochenberger2014,Glover2018}. Using an upper-triangular coefficient
convention, in which \(Q_{ji}=0\) for \(j>i\), a general QUBO problem can be
written as
\begin{equation}
    \min_{\mathbf{z}\in\{0,1\}^{N}}
    \mathbf{z}^{\top}Q\mathbf{z},
    \label{eq:general_qubo_matrix}
\end{equation}
where \(\mathbf{z}\) is a binary vector and \(Q\) is an upper-triangular
coefficient matrix. Under this convention, each off-diagonal coefficient
\(Q_{ij}\), \(i<j\), appears once in the corresponding quadratic polynomial.
Because \(z_i^2=z_i\) for binary variables, the objective is equivalently
written as
\begin{equation}
    \min_{\mathbf{z}\in\{0,1\}^{N}}
    \left(
        \sum_{i=1}^{N}Q_{ii}z_i
        +
        \sum_{i<j}Q_{ij}z_i z_j
    \right).
    \label{eq:general_qubo_expanded}
\end{equation}
Under the alternative symmetric-matrix convention, the corresponding
off-diagonal matrix entries would equal \(Q_{ij}/2\).

In the context of source-support selection, each binary variable \(z_i\)
indicates whether source-location group \(i\) is included in the active
support. The diagonal coefficients can encode candidate-specific selection
costs or data-fit contributions, whereas the off-diagonal coefficients can
encode pairwise couplings between selected candidates. Certain constraints
and prior preferences can be incorporated through quadratic penalty terms,
with auxiliary binary variables introduced when needed, while preserving the
QUBO form.

In the proposed method, target source-count information is represented using
the soft cardinality penalty
\begin{equation}
    \lambda
    \left(
        \sum_{i=1}^{N}z_i-k
    \right)^2,
    \qquad \lambda>0,
    \label{eq:general_soft_cardinality_penalty}
\end{equation}
which penalizes, but does not prohibit, deviations from \(k\). The specific
sensor-space construction of the diagonal and pairwise QUBO coefficients is
described in Section~\ref{sec:method}.

%% file: sections/03_method.tex
\section{Method}
\label{sec:method}

\subsection{Overview}
\label{subsec:method_overview}

Given prior information about a target source count \(k\), we formulate sparse
M/EEG source-support selection using a QUBO formulation with residual-aware
candidate screening over binary source-location-group selection variables.
Rather than optimizing the support directly over the full cortical source space,
the proposed method first constructs a reduced candidate set and candidate-specific
sensor-space templates, and then jointly selects the active support within that
candidate set.

The proposed method consists of four main steps. First, the sensor data and
gain matrix are preprocessed, and the cortical source space is represented
using source-location groups. Second, residual-aware screening is used to
construct a reduced candidate set. Third, a fixed sensor-space template is
constructed for each candidate using residual-based ridge fitting. Finally,
the candidate templates are selected jointly using a soft-cardinality QUBO
objective.

\subsection{Preprocessing and Source Grouping}
\label{subsec:preprocessing_source_grouping}

The sensor data and gain matrix were whitened using the sensor-noise covariance,
and a localization time window was selected around the evoked response of
interest following standard M/EEG inverse-modeling practice
\citep{Gramfort2014}.

The cortical source space was represented using groups defined at the
source-location level. In a fixed-orientation model, each source location
corresponds to one gain column. In a free-orientation model, the orientation
components associated with one cortical location form one gain block. The
binary variables therefore represent source-location groups rather than
individual gain columns.

Following standard depth-weighting practice, the whitened gain blocks were
normalized group-wise using a depth exponent of \(0.75\). The same normalized
gain blocks were used consistently for candidate screening, residual updating,
template construction, QUBO construction, and post-selection refitting. For a
gain matrix \(A\in\mathbb{R}^{m\times p}\), the ridge parameter was set to
\begin{equation*}
    \alpha(A)
    =
    \eta
    \max
    \left\{
        \frac{\operatorname{tr}(A^\top A)}{p},
        10^{-18}
    \right\},
\end{equation*}
where \(\eta\) is the relative ridge parameter. The resulting preprocessed
sensor data and source-group gain blocks are denoted by \(Y\) and \(G_i\),
respectively.

\subsection{Residual-Aware Candidate Screening}
\label{subsec:residual_candidate_screening}

Using all cortical source-location groups as QUBO variables would result in a
large optimization problem with a quadratic number of pairwise terms. We
therefore perform residual-aware screening to construct a reduced candidate
set \(\mathcal{C}\) with a predefined candidate budget.

For each source-location group \(i\), ridge-regularized fitting is performed
against a sensor-space residual \(R\):
\begin{equation}
    \widehat B_i(R)
    =
    \arg\min_B
    \left[
        \|R-G_iB\|_F^2
        +
        \alpha_i\|B\|_F^2
    \right],
    \qquad
    \widehat F_i(R)
    =
    G_i\widehat B_i(R),
    \label{eq:screening_group_fit}
\end{equation}
where \(\alpha_i=\alpha(G_i)\) follows the relative ridge rule defined above.
The screening score is based on the reduction in residual energy:
\begin{equation}
    s_i(R)
    =
    \sqrt{
        \max
        \left\{
            \|R\|_F^2
            -
            \|R-\widehat F_i(R)\|_F^2,
            0
        \right\}
    }.
    \label{eq:screening_score}
\end{equation}

Initial candidates are selected using \(s_i(Y)\). Starting from
\(R^{(0)}=Y\), residual-expansion round \(r\) ranks the source groups using
\begin{equation}
    \widetilde s_i^{(r)}
    =
    s_i\!\left(R^{(r-1)}\right)
    +
    w_{\mathrm{peak}}\,
    s_i\!\left(R_{:,t_{\mathrm{peak}}}^{(r-1)}\right),
    \label{eq:screening_combined_score}
\end{equation}
where \(t_{\mathrm{peak}}\) is the global-field-power peak within the selected
time window and \(w_{\mathrm{peak}}\) controls the contribution of the
peak-time score. High-ranking groups are added to the candidate pool.

The highest-ranking group not previously used for residual construction is
also added to a seed set \(\mathcal{S}^{(r)}\). The seed groups are jointly
refitted to the original preprocessed data, and the residual is updated as
\begin{align}
    \widehat B_{\mathcal{S}^{(r)}}
    &=
    \arg\min_B
    \left[
        \|Y-G_{\mathcal{S}^{(r)}}B\|_F^2
        +
        \alpha_{\mathcal{S}^{(r)}}\|B\|_F^2
    \right],
    \nonumber\\
    R^{(r)}
    &=
    Y
    -
    G_{\mathcal{S}^{(r)}}
    \widehat B_{\mathcal{S}^{(r)}}.
    \label{eq:screening_residual_update}
\end{align}

Within each ranking list, retained candidates were assigned a normalized
rank-based priority, with higher-ranked candidates receiving higher priority.
For candidates appearing in multiple lists, the maximum priority was retained.
The candidate union was then ordered by decreasing priority and, for equal
priority, by decreasing initial score \(s_i(Y)\). Candidates were retained up
to the predefined budget; if the union was smaller than the budget, additional
candidates were taken from the initial global ranking. This produced the final
candidate set \(\mathcal{C}\). Final support selection was subsequently
performed over \(\mathcal{C}\) using the soft-cardinality QUBO.

\subsection{Residual-Aware Sensor-Space Template Construction}
\label{subsec:residual_template_construction}

After candidate screening, one fixed sensor-space template is constructed for
each candidate group \(i\in\mathcal{C}\). The templates are estimated from a
sequence of successively updated residuals to reduce repeated assignment of
dominant signal components to multiple candidates. The initial residual is
defined as
\begin{equation}
    \widetilde R^{(0)}
    =
    Y.
    \label{eq:template_initial_residual}
\end{equation}

Template refinement is performed for
\begin{equation}
    n_{\mathrm{temp}}
    =
    \min\{k,|\mathcal{C}|\}
    \label{eq:template_round_count}
\end{equation}
rounds. Thus, the target source-count information determines the number of
residual-deflation steps used during template construction, whereas the final
support cardinality is determined by the subsequent soft-cardinality QUBO.

At each refinement round, all candidates are fitted to the current residual
and ranked according to their reduction in residual energy. Among the first
\(M_{\mathrm{temp}}\) candidates in the overall ranking, candidates whose
templates have not yet been assigned retain the fit obtained from the current
residual. Let \(r_i\) denote the round at which candidate \(i\) is assigned.
Its coefficient matrix is
\begin{equation}
    \widehat B_i
    =
    \arg\min_B
    \left[
        \|\widetilde R^{(r_i)}-G_iB\|_F^2
        +
        \alpha_i\|B\|_F^2
    \right],
    \label{eq:template_candidate_fit}
\end{equation}
and its fixed sensor-space template is
\begin{equation}
    \widehat F_i
    =
    G_i\widehat B_i.
    \label{eq:template_fixed_contribution}
\end{equation}

The highest-ranking candidate not previously used for residual construction
is also added to a template seed set. The accumulated seed groups are jointly
refitted to the original data \(Y\), and the resulting residual is used in the
next refinement round. Candidates whose templates have not been assigned
during the refinement rounds are fitted using the final residual. Consequently,
every candidate \(i\in\mathcal{C}\) is associated with one fixed sensor-space
template \(\widehat F_i\).

The binary variables introduced below control the inclusion or exclusion of
these fixed templates rather than the continuous coefficient matrices
themselves. The resulting QUBO objective is therefore a surrogate constructed
from fixed candidate-wise sensor-space contributions and is not equivalent to
jointly refitting the continuous coefficients for every possible binary
support.

\subsection{Soft-Cardinality QUBO Support Selection}
\label{subsec:soft_cardinality_qubo}

For each candidate group \(i\in\mathcal{C}\), let \(z_i\in\{0,1\}\) indicate
whether that group is selected. The sensor-space signal predicted by a binary
selection vector \(z\) is
\begin{equation}
    \widehat Y(z)
    =
    \sum_{i\in\mathcal{C}}
    z_i\widehat F_i.
    \label{eq:qubo_predicted_signal}
\end{equation}

The primary source-support selection problem is formulated as
\begin{equation}
    \min_{z\in\{0,1\}^{|\mathcal{C}|}}
    \left\|
        Y
        -
        \sum_{i\in\mathcal{C}}
        z_i\widehat F_i
    \right\|_F^2
    +
    \lambda
    \left(
        \sum_{i\in\mathcal{C}}z_i-k
    \right)^2.
    \label{eq:soft_cardinality_qubo_objective}
\end{equation}

The target source count \(k\) is not imposed as a feasibility constraint.
Instead, deviations from \(k\) are penalized within the unconstrained binary
objective. The estimated support cardinality
\begin{equation}
    \widehat k
    =
    \sum_{i\in\mathcal{C}}z_i
    \label{eq:estimated_support_cardinality}
\end{equation}
is therefore determined by the optimized binary solution and may differ from
\(k\).

Ignoring terms that are constant with respect to \(z\),
Equation~\eqref{eq:soft_cardinality_qubo_objective} can be written as
\begin{equation}
    \sum_{i\in\mathcal{C}}a_i z_i
    +
    \sum_{\substack{i,j\in\mathcal{C}\\i<j}}
    b_{ij}z_i z_j,
    \label{eq:soft_cardinality_qubo_polynomial}
\end{equation}
where
\begin{equation}
    a_i
    =
    \|\widehat F_i\|_F^2
    -
    2\langle Y,\widehat F_i\rangle_F
    +
    \lambda(1-2k),
    \label{eq:soft_cardinality_linear_coefficient}
\end{equation}
and
\begin{equation}
    b_{ij}
    =
    2\langle\widehat F_i,\widehat F_j\rangle_F
    +
    2\lambda.
    \label{eq:soft_cardinality_pair_coefficient}
\end{equation}

The linear coefficients combine the candidate-specific data-fit contribution
with the cardinality bias. The pairwise coefficients represent sensor-space
overlap between fixed candidate templates together with the quadratic coupling
induced by the soft-cardinality penalty. They should not be interpreted as
physiological interactions between cortical sources.

To scale the cardinality penalty relative to the candidate-wise data-fit terms,
define
\begin{equation}
    q_i^{(0)}
    =
    \|\widehat F_i\|_F^2
    -
    2\langle Y,\widehat F_i\rangle_F.
    \label{eq:unpenalized_qubo_diagonal}
\end{equation}
The cardinality penalty is set as
\begin{equation}
    \lambda
    =
    c_{\mathrm{card}}
    \operatorname{median}_{i\in\mathcal{C}}
    \left[
    \max\left\{
    \left|q_i^{(0)}\right|,
    10^{-18}
    \right\}
    \right],
    \label{eq:cardinality_penalty_scaling}
\end{equation}
where \(c_{\mathrm{card}}>0\) is a dimensionless penalty multiplier. In the
implementation, a numerical floor of \(10^{-18}\) was applied to
\(\lvert q_i^{(0)}\rvert\). The experimental value of
\(c_{\mathrm{card}}\) and the QUBO solver settings are reported in
Section~\ref{sec:evaluation}.

The soft-cardinality QUBO is minimized over the unconstrained binary domain
using classical simulated annealing, and the lowest-energy sample is retained.
No post hoc projection or truncation is applied to force the selected
cardinality to equal \(k\).

For sensor-space reconstruction and residual calculation, the continuous
coefficients are jointly refitted on the source groups selected by the QUBO.
This post-selection refit does not alter the selected binary support.

%% file: sections/04_evaluation.tex
\section{Evaluation}
\label{sec:evaluation}

We evaluate the proposed QUBO formulation with residual-aware candidate
screening using synthetic MEG simulations with known active-source locations.
The evaluation focuses on source localization rather than waveform reconstruction
or amplitude estimation. Within each condition, all methods use the same
simulated data, forward model, noise covariance, source space, and localization
window.

\subsection{Synthetic MEG Data}
\label{subsec:synthetic_meg_data}

Synthetic MEG data are generated using the cortical source space, MEG forward
model, raw recording, and noise covariance of the MNE-Python sample dataset
\citep{Gramfort2013,Gramfort2014}. Although the proposed formulation applies
to the general M/EEG inverse problem, the present experiments are restricted
to MEG simulations.

True sources are sampled from a visibility-constrained subset defined using
the noise-whitened, fixed-orientation lead field. Locations are retained when
their column-norm visibility score is at or above the 45th percentile and
their radial distance from the median source-space coordinate is at or above
the 25th percentile. For the lead field used here, the Frobenius-norm and
maximum-singular-value criteria reduce to this same column-norm score. These
criteria are used only for ground-truth sampling.

Within this eligible pool, source locations are sampled pseudorandomly.
Simultaneously active sources are required to be separated by at least
\(30\,\mathrm{mm}\) in three-dimensional Euclidean distance. Sensor-level
responses are generated using fixed surface-normal source orientations.

Each active source is assigned a Gaussian-windowed \(30\,\mathrm{Hz}\)
sinusoidal waveform,
\begin{equation}
    x_i(t)
    =
    A
    \sin(2\pi f t)
    \exp
    \left[
        -\frac{1}{2}
        \left(
            \frac{t-\tau_i}{\sigma}
        \right)^2
    \right],
    \label{eq:synthetic_source_waveform}
\end{equation}
where \(A=50\,\mathrm{nA\,m}\), \(f=30\,\mathrm{Hz}\), and
\(\sigma=50\,\mathrm{ms}\) is the standard deviation of the Gaussian envelope.
The source-specific peak time is
\begin{equation}
    \tau_i
    =
    150\,\mathrm{ms}
    +
    \epsilon_i,
    \qquad
    \epsilon_i
    \sim
    \mathcal{N}
    \left(
        0,
        (50\,\mathrm{ms})^2
    \right).
    \label{eq:synthetic_peak_time_jitter}
\end{equation}
The simulated time axis contains 300 samples beginning at
\(-100\,\mathrm{ms}\), using the sampling frequency of the sample dataset.

Colored sensor noise is generated from the sample-dataset noise covariance
using a fifth-order IIR noise model fitted to the \(60\)--\(180\,\mathrm{s}\)
segment of the sample raw recording. The effective noise level is controlled
through the number of averaged epochs, \(n_{\mathrm{ave}}\), in the evoked-data
simulation, with larger values corresponding to lower noise levels. The main
benchmark uses two active sources and \(n_{\mathrm{ave}}=30\).

For each trial, the trial index seeds source-location sampling, waveform
peak-time jitter, and simulated annealing. The sensor-noise generator is not
explicitly seeded, so exact noise realizations are not guaranteed to be
reproducible. Localization is performed within an approximately
\(80\,\mathrm{ms}\) window centered on the global-field-power peak of the
whitened sensor data.

\subsection{Compared Methods}
\label{subsec:compared_methods}

The proposed method, denoted by QUBO, refers to the soft-\(k\) formulation in
Section~\ref{subsec:soft_cardinality_qubo}. The true source count \(k\) enters
through a quadratic cardinality penalty rather than a hard constraint, and the
source groups returned by the binary optimization are used directly without
forcing their number to equal \(k\).

QUBO is compared with minimum-norm estimation (MNE), dynamic statistical
parametric mapping (dSPM), mixed-norm estimation (MxNE), RAP-MUSIC, and the
linearly constrained minimum-variance beamformer (LCMV)
\citep{Hamalainen1994,Dale2000,Gramfort2012,Mosher1999,VanVeen1997}.

For methods producing distributed source maps, each source location is scored
using its maximum absolute activity over the same localization window used by
QUBO. Vertices above the 97.5th percentile of the source-score distribution
are grouped into connected spatial components using a \(15\,\mathrm{mm}\)
three-dimensional Euclidean linkage distance. The highest-scoring vertex in
each component is treated as its cluster peak, and up to \(k\) highest-scoring
cluster peaks are retained. Fewer than \(k\) locations may be returned when
the estimated source map does not contain enough distinct positive-scoring
components. RAP-MUSIC dipole estimates are mapped to their nearest cortical
source locations in three-dimensional Euclidean distance.

MNE and dSPM use an SNR of \(3\), corresponding to
\(\lambda^2_{\mathrm{MNE}}=1/9\), and a depth-weighting exponent of \(0.8\).
MxNE uses \(\alpha=70\) on MNE-Python's \([0,100)\) regularization scale,
a depth-weighting exponent of \(0.8\), coefficient debiasing, and time PCA.
RAP-MUSIC is requested to estimate \(k\) dipoles.

LCMV uses a regularization parameter of \(0.05\), max-power orientation,
unit-noise-gain normalization, a depth-weighting exponent of \(0.8\), and
matrix inversion. Its data covariance is estimated empirically from the
preprocessing-selected localization window, and the simulation noise
covariance is supplied during filter construction. Rank handling is left to
MNE-Python's default estimation.

The same \(k\) is used as the target count in QUBO, the requested dipole count
in RAP-MUSIC, and the maximum number of retained peaks for distributed source
maps.

\subsection{Localization Metrics}
\label{subsec:evaluation_metrics}

True and estimated source locations are matched one-to-one by minimizing the
total three-dimensional Euclidean distance. Let $\mathcal{M}$ denote the
resulting set of $\min(k,\widehat{k})$ matched pairs, where $k$ and
$\widehat{k}$ are the numbers of true and estimated sources, respectively.
The primary localization error is defined as
\begin{equation}
E_{\mathrm{loc}}
=
\frac{
\displaystyle
\sum_{(p,q)\in\mathcal{M}}
\left\|\mathbf{r}_p-\hat{\mathbf{r}}_q\right\|_2
+
D_{\mathrm{card}}\left|\hat{k}-k\right|
}{k},
\label{eq:localization_error}
\end{equation}
where $\mathbf{r}_p$ and $\widehat{\mathbf{r}}_q$ are the true and estimated
source coordinates, respectively, and $k$ and $\widehat{k}$ denote the true and
estimated source counts. We set $D_{\mathrm{card}} = 50\,\mathrm{mm}$ per unit
of source-count mismatch. Thus, the term $D_{\mathrm{card}} \lvert \widehat{k} - k \rvert$
symmetrically penalizes both missing and additional estimated sources, treating
each source-count mismatch as a large localization failure. Consequently, comparisons
based on $E_{\mathrm{loc}}$ are conditional on the chosen cardinality-mismatch penalty.

For a distance threshold $d$, the true-source-normalized hit rate is defined as
\begin{equation}
\operatorname{Hit}@d
=
\frac{1}{k}
\sum_{(p,q)\in\mathcal{M}}
\mathbb{I}
\left[
\left\lVert
\mathbf{r}_p-\widehat{\mathbf{r}}_q
\right\rVert_2
\le d
\right],
\label{eq:true_source_normalized_hit_rate}
\end{equation}
where $\mathbb{I}[\cdot]$ is the indicator function and unmatched true sources
count as misses. We report trial-averaged Hit@10 and Hit@20 using
$d=10\,\mathrm{mm}$ and $d=20\,\mathrm{mm}$, respectively.

We additionally report the mean estimated source count and the exact-count,
under-selection, and over-selection rates. Localization error is summarized
using the mean, standard deviation, median, and interquartile range and is
displayed using box plots with individual trial points.

\subsection{Benchmark Settings}
\label{subsec:benchmark_settings}

The main benchmark comprises 100 simulated trials with \(k=2\) and
\(n_{\mathrm{ave}}=30\), comparing QUBO with MNE, dSPM, MxNE, RAP-MUSIC,
and LCMV.

The QUBO uses a candidate budget of 300. Candidate screening consists of one
initial global screening step followed by four residual-expansion rounds. The
initial screening retains the 158 highest-ranking groups, and up to 120 groups
are added from each residual-based ranking. The candidate union is subsequently
ordered and truncated or supplemented to obtain the final 300-candidate set.
The peak-time score weight is \(0.5\), the relative ridge parameter is
\(\eta=0.01\), and up to 120 previously unassigned templates are retained per
template-refinement round.

The cardinality-penalty multiplier is \(c_{\mathrm{card}}=1.5\). The
soft-\(k\) QUBO is solved using the D-Wave \texttt{neal} classical simulated
annealer with 1000 reads and 1000 sweeps. The trial index is used as the
annealing random state, and the lowest-energy sample is retained without
cardinality projection.

For the main benchmark, exact candidate coverage is also evaluated to examine
the effect of candidate screening on the final QUBO localization performance.
It is defined as
\begin{equation}
    C_{\mathrm{exact}}
    =
    \frac{
        \left|
            \mathcal{T}\cap\mathcal{C}
        \right|
    }{
        |\mathcal{T}|
    },
    \label{eq:exact_candidate_coverage}
\end{equation}
where \(\mathcal{T}\) is the set of true source indices and
\(\mathcal{C}\) is the screened candidate set. Exact candidate coverage is
computed only as a post hoc diagnostic and is not used during candidate
screening, template construction, or QUBO optimization.

To evaluate robustness to sensor noise, the number of active sources is fixed
at \(k=2\), and the number of averaged epochs is varied over
\begin{equation}
    n_{\mathrm{ave}}
    \in
    \{5,10,20,30,50,100\}.
    \label{eq:noise_analysis_nave_values}
\end{equation}
Each noise condition comprises 100 simulated trials.

To evaluate the effect of the number of active sources,
\(n_{\mathrm{ave}}=30\) is fixed, and the source count is varied over
\begin{equation}
    k
    \in
    \{1,2,3\}.
    \label{eq:source_count_analysis_values}
\end{equation}
Each source-count condition comprises 100 simulated trials.

The cardinality ablation compares soft-\(k\) QUBO with an exact-\(k\) reference
and a sparsity-penalized no-\(k\) variant under \(k=2\) and
\(n_{\mathrm{ave}}=30\). Here, no-\(k\) denotes removal of the target-cardinality
term only from the final QUBO objective. All variants use the same 100 trials,
candidate sets, and fixed sensor-space templates.

The exact-\(k\) reference solves
\begin{equation}
    \begin{aligned}
        \min_{z\in\{0,1\}^{|\mathcal{C}|}}
        \quad&
        \left\|
            Y
            -
            \sum_{i\in\mathcal{C}}
            z_i\widehat F_i
        \right\|_F^2
        \\
        \text{subject to}
        \quad&
        \sum_{i\in\mathcal{C}}z_i=k.
    \end{aligned}
    \label{eq:exact_k_reference_objective}
\end{equation}
Under the \(k=2\) ablation condition, it is solved by vectorized enumeration
of all \(\binom{|\mathcal{C}|}{2}\) feasible candidate supports. For a
300-candidate set, this corresponds to 44,850 supports.

The no-\(k\) formulation solves
\begin{equation}
    \min_{z\in\{0,1\}^{|\mathcal{C}|}}
    \left\|
        Y
        -
        \sum_{i\in\mathcal{C}}
        z_i\widehat F_i
    \right\|_F^2
    +
    \gamma
    \sum_{i\in\mathcal{C}}z_i,
    \label{eq:no_cardinality_sparsity_objective}
\end{equation}
where
\begin{equation}
    \gamma
    =
    c_{\mathrm{sp}}
    \operatorname{median}_{i\in\mathcal{C}}
    \left[
        \max
        \left\{
            |q_i^{(0)}|,
            10^{-18}
        \right\}
    \right].
    \label{eq:no_cardinality_sparsity_penalty_scaling}
\end{equation}
The implementation uses \(c_{\mathrm{sp}}=10.0\). Soft-\(k\) and no-\(k\) use
the same simulated-annealing settings, whereas exact-\(k\) is solved by
enumeration. Because upstream template construction still uses \(k\), no-\(k\)
is not a fully \(k\)-independent pipeline. This ablation therefore compares
localization and source-count selection rather than solver runtime.

Unless otherwise stated, the candidate-screening, template-construction, and
simulated-annealing settings described above are retained across the main
benchmark, noise analysis, source-count analysis, and cardinality ablation.
These four analyses use separately generated simulation sets. Within each
experimental condition, all compared methods or QUBO variants are evaluated
on the same simulated trials.

%% file: sections/05_results.tex
\section{Results}
\label{sec:results}

We present the performance of the proposed QUBO-based source-support
selection method using the benchmark settings defined in
Section~\ref{subsec:benchmark_settings}. We first compare QUBO with the
baseline source-localization methods under the default simulation setting
and examine how candidate coverage affects the final QUBO localization
error. We then evaluate robustness to sensor noise and the effect of the
number of active sources. Finally, we compare different forms of
source-cardinality control.

\subsection{Main Benchmark Performance}
\label{subsec:main_benchmark_results}

\begin{figure}[H]
    \centering
    \includegraphics[width=0.92\linewidth]
    {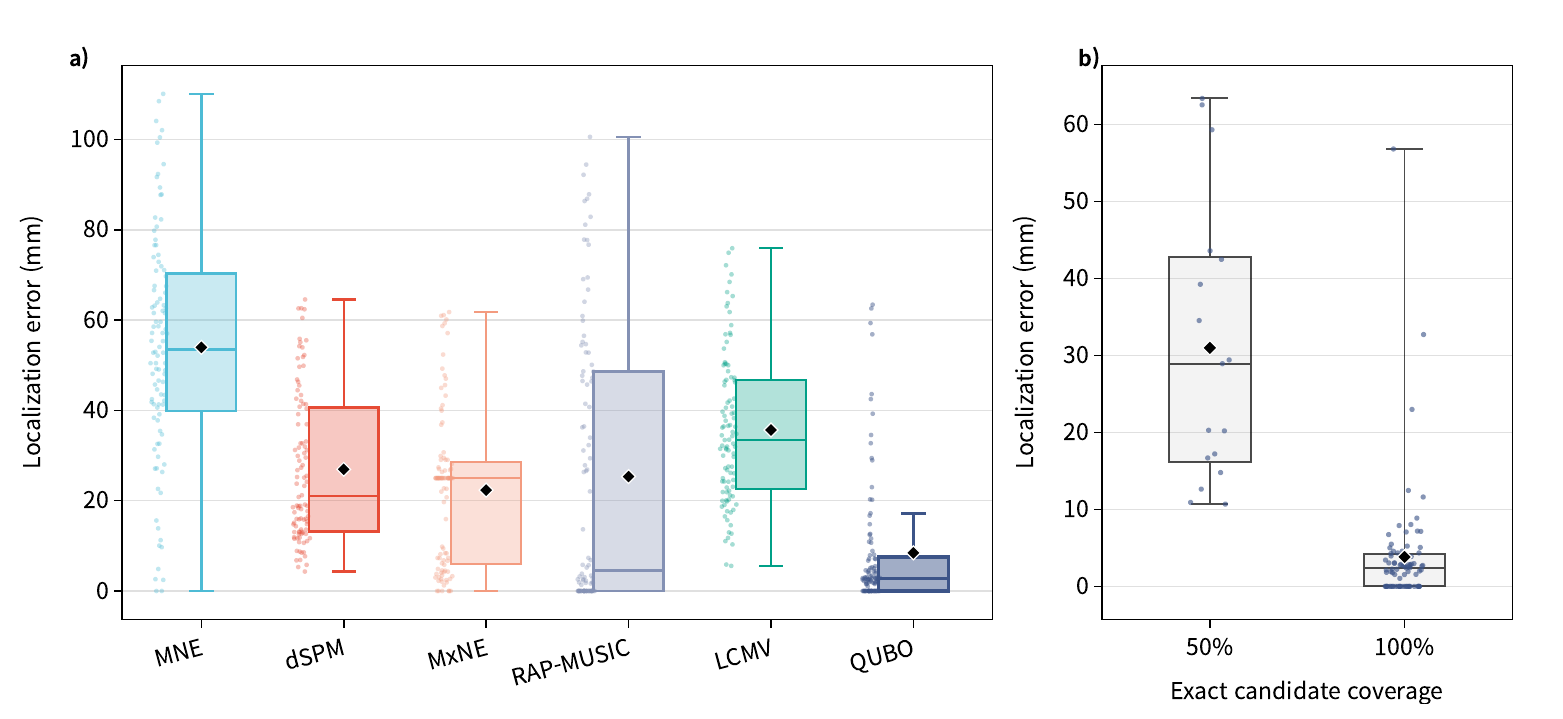}
    \caption{
    Main benchmark and candidate-coverage analysis under the
    \(k=2\), \(n_{\mathrm{ave}}=30\) condition.
    (a) True-source-normalized localization error distributions for QUBO
    and the baseline source-localization methods. Each point represents one
    simulated trial, the box plots summarize the trial-level distributions,
    and the diamond markers indicate the means.
    (b) QUBO true-source-normalized localization error grouped by exact
    candidate coverage. Exact candidate coverage is the fraction of true
    source indices included in the screened candidate set before simulated
    annealing.
    }
    \label{fig:main_benchmark}
\end{figure}

Figure~\ref{fig:main_benchmark}a shows the true-source-normalized
localization error distributions of QUBO and the baseline methods over
100 trials with \(k=2\) and \(n_{\mathrm{ave}}=30\). Each point represents
one simulated trial, the box plots summarize the trial-level distributions,
and the diamond markers indicate the means.

QUBO produced the lowest errors overall under this metric, with most trials
concentrated near zero. MxNE showed the lowest mean error among the baseline
methods. RAP-MUSIC also showed a low median error, but its distribution was
considerably wider. MNE produced the largest errors among the evaluated
methods.

The quantitative results are summarized in
Table~\ref{tab:main_benchmark}. QUBO achieved the lowest mean
true-source-normalized localization error,
\(8.45 \pm 14.37\,\mathrm{mm}\). The best-performing baseline according
to this metric was MxNE, with \(22.35 \pm 16.99\,\mathrm{mm}\).
This corresponds to a \(62.2\%\) reduction in mean error relative to MxNE
under the evaluated metric.

QUBO also achieved the lowest median error,
\(2.76\,\mathrm{mm}\), with an interquartile range of
\([0.00,7.37]\,\mathrm{mm}\). MxNE had a median error of
\(25.00\,\mathrm{mm}\), with an interquartile range of
\([6.06,28.37]\,\mathrm{mm}\). RAP-MUSIC had a low median error of
\(4.51\,\mathrm{mm}\), but its interquartile range was substantially wider,
\([0.00,48.61]\,\mathrm{mm}\). These results indicate that QUBO achieved
both a lower mean error and a more concentrated trial-level error
distribution under the default simulation condition.

QUBO also achieved the highest true-source-normalized hit rates, with
Hit@10 and Hit@20 values of \(84.5\%\) and \(90.0\%\), respectively.
Among the baseline methods, MxNE achieved the highest Hit@10 value of
\(64.5\%\), whereas dSPM and MxNE both achieved the highest baseline
Hit@20 value of \(68.0\%\).

\input{tables/table01_main_benchmark_summary_full.tex}

QUBO returned two source locations in all 100 trials, matching the target
source count. MNE, dSPM, RAP-MUSIC, and LCMV also returned two locations
in every trial. MxNE returned two locations in 65 trials and one location
in the remaining 35 trials. The resulting source-count mismatch was included
in the localization error using the symmetric cardinality-mismatch term
$D_{\mathrm{card}}\lvert\widehat{k}-k\rvert$, with
$D_{\mathrm{card}}=50\,\mathrm{mm}$. In each of the 35 trials with only one
returned location, $\lvert\widehat{k}-k\rvert=1$, corresponding to a
$25\,\mathrm{mm}$ contribution to $E_{\mathrm{loc}}$ after normalization by
the true source count $k=2$. No over-selection occurred in the main benchmark.

\FloatBarrier

\subsection{Candidate Coverage and QUBO Performance}
\label{subsec:candidate_coverage_results}

Figure~\ref{fig:main_benchmark}b groups the QUBO localization errors
according to exact candidate coverage. Exact candidate coverage is the
fraction of true source indices included in the screened candidate set before
simulated annealing. Because the main benchmark contains two true sources,
\(50\%\) coverage indicates that one true source was included, whereas
\(100\%\) coverage indicates that both true sources were included.

Both true sources were included in the screened candidate set in 83 of the
100 trials. The mean localization error in these trials was
\(3.83\,\mathrm{mm}\). In the remaining 17 trials, only one of the two
true sources was included, and the mean localization error increased to
\(31.00\,\mathrm{mm}\).

Simulated annealing selects source groups only from the screened candidate
set. Therefore, a true source omitted during candidate screening cannot be
selected at its exact location during the subsequent binary optimization,
although a nearby candidate may still provide an approximate estimate. The
large difference between the two coverage groups indicates that candidate
screening was a major limiting factor in the high-error trials.
\FloatBarrier

\subsection{Robustness to Sensor Noise}
\label{subsec:noise_results}

\begin{figure}[H]
    \centering
    \includegraphics[width=\linewidth]
    {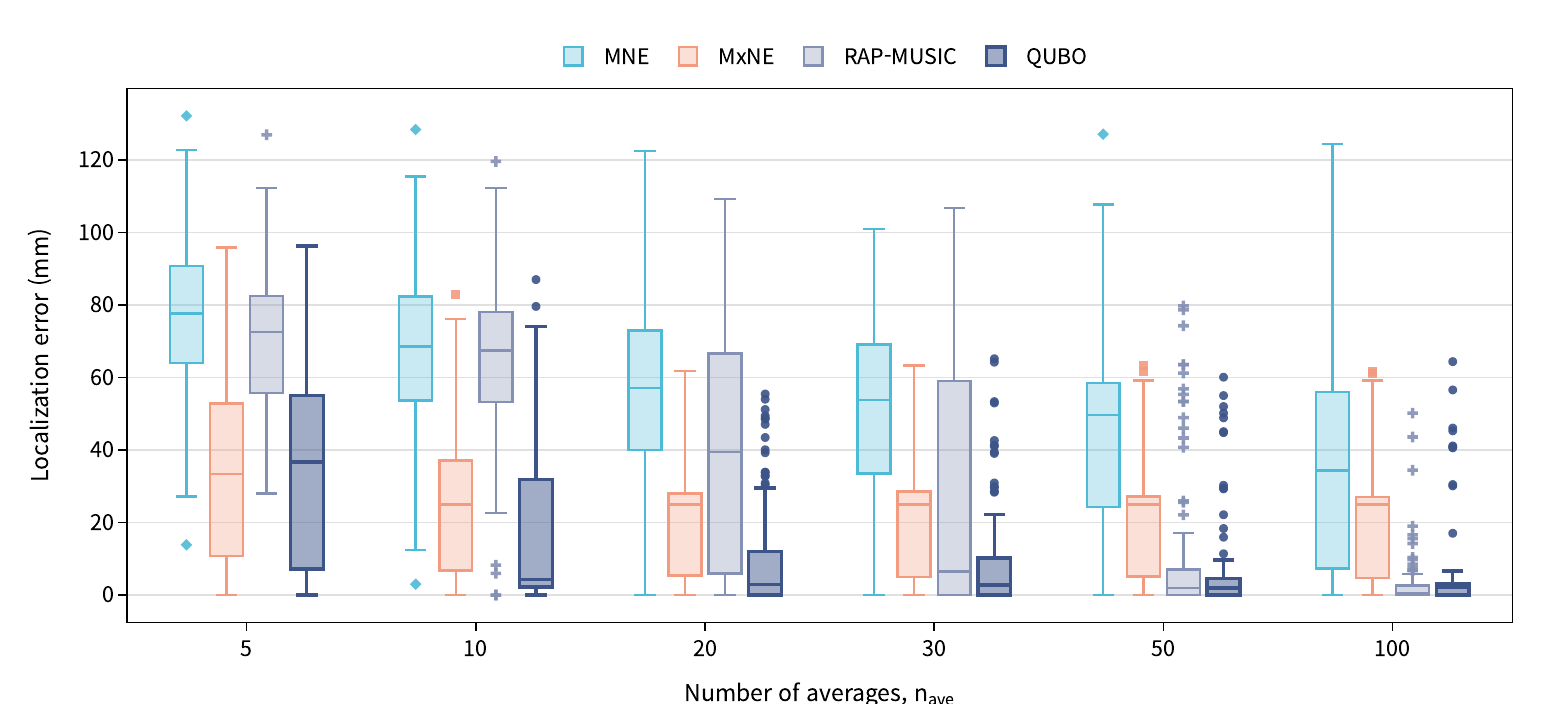}
    \caption{
        Localization-error distributions across sensor-noise conditions.
        Results are shown as a function of the number of averages,
        \(n_{\mathrm{ave}}\), with larger values corresponding to lower effective
        sensor noise. Box plots summarize the trial-level distributions for QUBO,
        MNE, MxNE, and RAP-MUSIC.
    }
    \label{fig:noise_robustness}
\end{figure}

Figure~\ref{fig:noise_robustness} shows the localization-error distributions
across the tested sensor-noise conditions. As \(n_{\mathrm{ave}}\) increased,
the QUBO error distribution generally shifted toward lower values.

QUBO performance improved substantially as the effective noise level
decreased. Its mean localization error decreased from
\(34.76\,\mathrm{mm}\) at \(n_{\mathrm{ave}}=5\) to
\(17.38\,\mathrm{mm}\) at \(n_{\mathrm{ave}}=10\) and
\(10.74\,\mathrm{mm}\) at \(n_{\mathrm{ave}}=20\). The mean error further
decreased to \(9.19\,\mathrm{mm}\) at \(n_{\mathrm{ave}}=30\),
\(6.76\,\mathrm{mm}\) at \(n_{\mathrm{ave}}=50\), and
\(5.61\,\mathrm{mm}\) at \(n_{\mathrm{ave}}=100\).

At the highest-noise condition, \(n_{\mathrm{ave}}=5\), QUBO and MxNE
showed similar performance. QUBO had a slightly lower mean error
(\(34.76\) versus \(36.08\,\mathrm{mm}\)), whereas MxNE had a slightly
lower median error (\(33.44\) versus \(36.71\,\mathrm{mm}\)).
At lower noise levels, QUBO consistently had lower mean localization errors
than MxNE.

Among the four methods shown in Figure~\ref{fig:noise_robustness}, QUBO
achieved the lowest mean localization error at
\(n_{\mathrm{ave}}=5,10,20,30,\) and \(50\). RAP-MUSIC showed relatively
large errors under high-noise conditions but improved markedly as the noise
level decreased. At \(n_{\mathrm{ave}}=50\), QUBO and RAP-MUSIC had nearly
identical median errors (\(1.93\) and \(1.90\,\mathrm{mm}\), respectively),
although QUBO had the lower mean error
(\(6.76\) versus \(10.97\,\mathrm{mm}\)). At
\(n_{\mathrm{ave}}=100\), RAP-MUSIC achieved the lowest mean localization
error among the four methods, with \(3.13\,\mathrm{mm}\), compared with
\(5.61\,\mathrm{mm}\) for QUBO.

The hit-rate results showed a similar improvement as the effective noise
level decreased. For QUBO, Hit@10 increased from \(48.0\%\) at
\(n_{\mathrm{ave}}=5\) to \(92.5\%\) at
\(n_{\mathrm{ave}}=100\), while Hit@20 increased from \(57.0\%\) to
\(95.0\%\). At \(n_{\mathrm{ave}}=100\), RAP-MUSIC achieved slightly higher
Hit@10 and Hit@20 values of \(94.5\%\) and \(96.5\%\), respectively.

Overall, QUBO achieved the lowest mean localization error among the four
methods shown in five of the six tested noise conditions. RAP-MUSIC achieved
a lower mean error than QUBO only at the lowest-noise condition.

\FloatBarrier

\subsection{Robustness to the Number of Active Sources}
\label{subsec:source_count_results}

\begin{figure}[H]
    \centering
    \includegraphics[width=\linewidth]{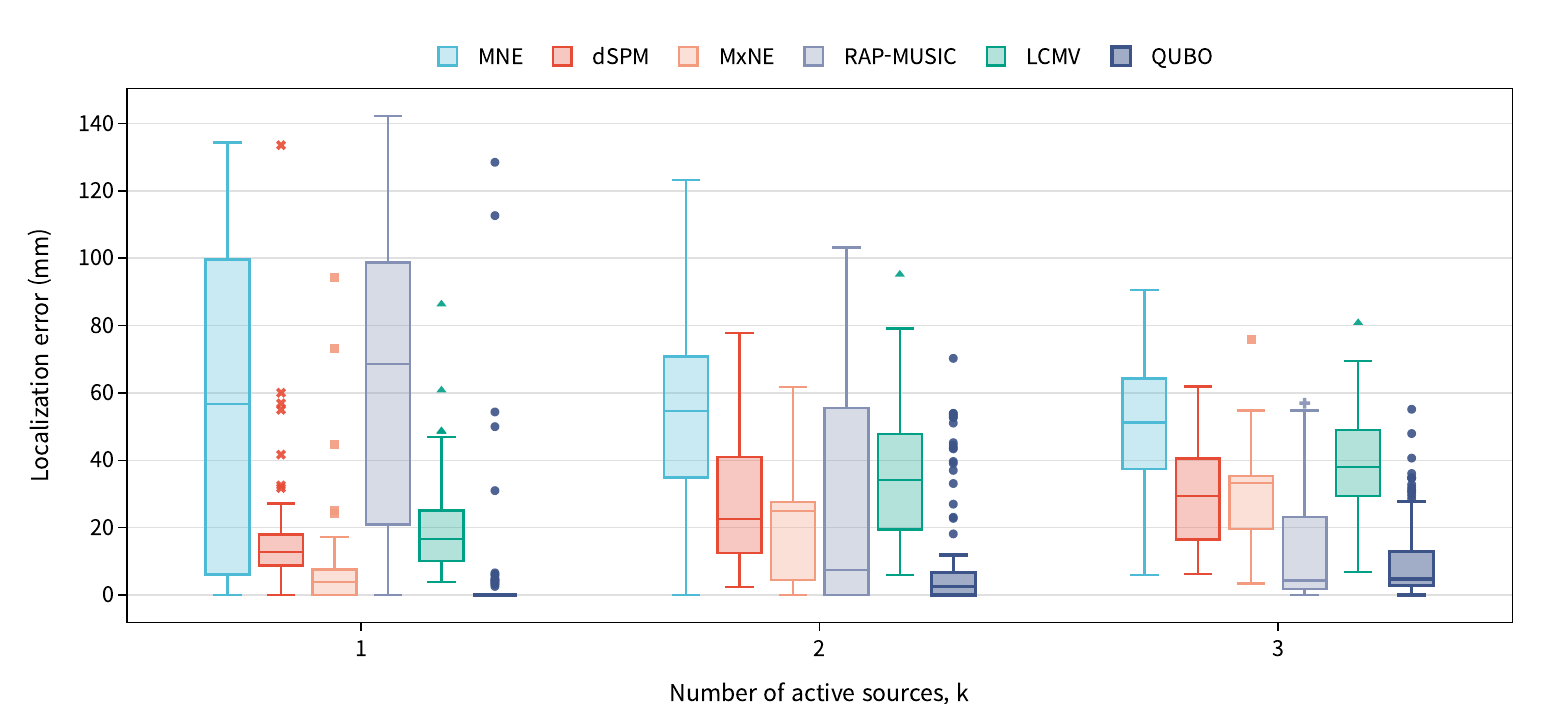}
    \caption{
        Robustness to the number of active sources. Localization error distributions
        are compared for \(k=1\), \(k=2\), and \(k=3\). Box plots summarize the
        trial-level error distributions for each method, with outlier points shown
        separately.
    }
    \label{fig:source_count_robustness}
\end{figure}

Figure~\ref{fig:source_count_robustness} shows how the trial-level localization
error distributions changed with the number of active sources. We compared
\(k=1\), \(k=2\), and \(k=3\) to evaluate whether the proposed binary
source-selection formulation remains effective beyond the default \(k=2\)
setting.

Across the tested source-count settings, QUBO generally showed
localization-error distributions concentrated closer to zero than those of
most baseline methods. For \(k=1\), the QUBO distribution had a median error
of \(0.00\) mm and an interquartile range of \([0.00, 0.00]\) mm, indicating
exact recovery in at least half of the trials. MxNE also showed relatively low
errors in this setting, whereas MNE and RAP-MUSIC had broader error
distributions.

For the default \(k=2\) setting, QUBO again showed a low error distribution,
with a median error of \(2.56\) mm and an interquartile range of
\([0.00, 6.60]\) mm. RAP-MUSIC had the next-lowest median error of
\(7.49\) mm, although its interquartile range was substantially wider
\([0.00, 54.77]\) mm. MxNE had a median error of \(25.00\) mm and an
interquartile range of \([4.53, 27.49]\) mm. These \(k=2\) results were
obtained from the source-count robustness benchmark and are therefore
reported separately from the main benchmark.

For \(k=3\), localization became more difficult, and the QUBO distribution
shifted toward larger errors. The median QUBO error increased to \(4.74\) mm,
with an interquartile range of \([2.84, 12.77]\) mm. RAP-MUSIC achieved a
slightly lower median error of \(4.37\) mm but exhibited a wider
interquartile range of \([1.87, 23.13]\) mm. Thus, QUBO and RAP-MUSIC showed
comparable central tendencies in the three-source setting, while QUBO
produced a more concentrated error distribution. MxNE, dSPM, LCMV, and MNE
showed higher median errors or broader error distributions.

Overall, these results indicate that the proposed QUBO formulation maintained
competitive localization performance beyond the default \(k=2\) setting
within the tested source-count range. However, the shift in the QUBO error
distribution as \(k\) increased suggests that support selection becomes more
challenging as the number of simultaneously active sources grows. For larger
\(k\), the stability of candidate screening and cardinality control is likely
to become increasingly important.

\FloatBarrier

\subsection{Cardinality Ablation}
\label{subsec:cardinality_ablation_results}

\begin{figure}[H]
    \centering
    \includegraphics[width=\linewidth]
    {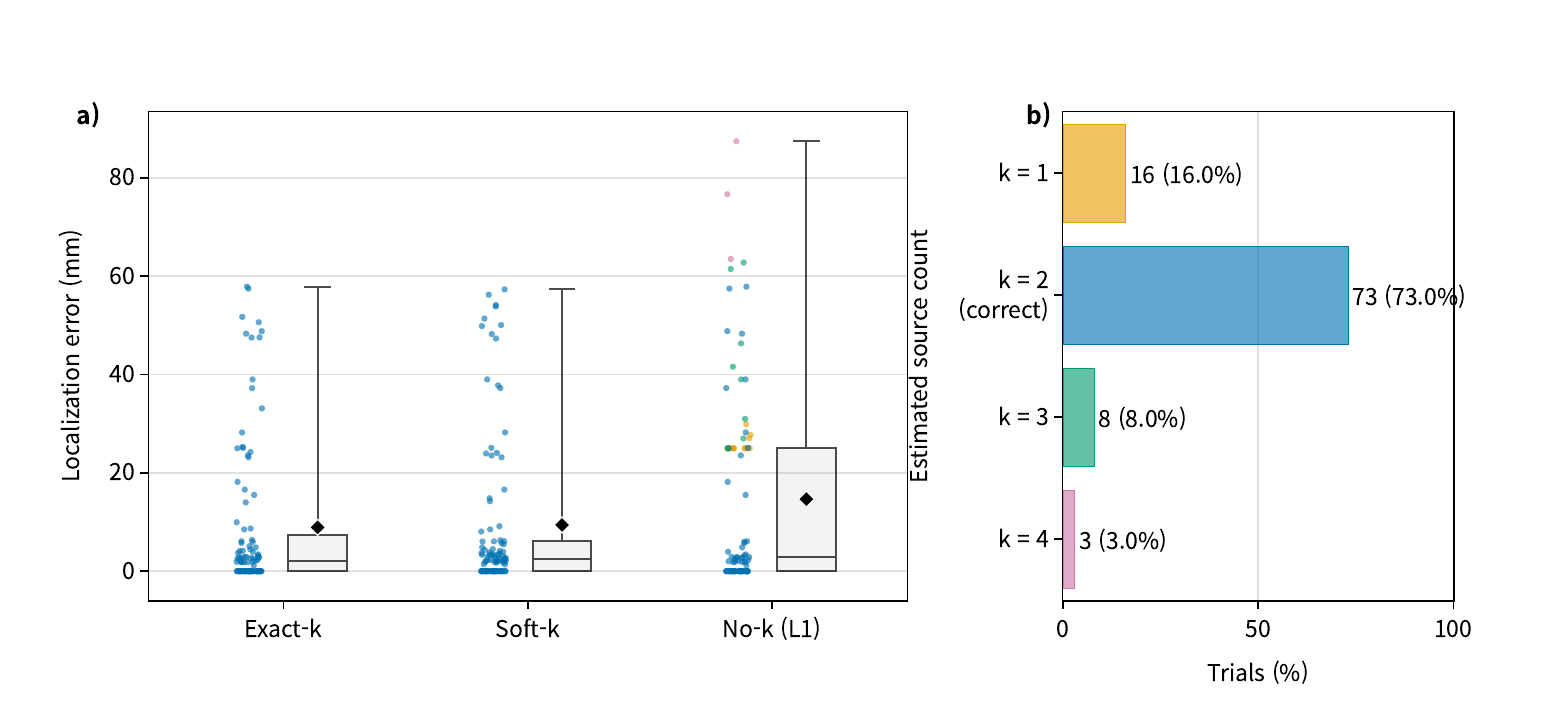}
    \caption{
    Cardinality ablation of the QUBO source-selection formulation.
    (a) Localization error for the exact-\(k\), soft-\(k\), and
    sparsity-penalized no-\(k\) variants. Localization error includes the
    symmetric source-count mismatch penalty with
    \(D_{\mathrm{card}}=50\,\mathrm{mm}\). Dots represent individual trials
    and are colored by the estimated source count \(\hat{k}\); gray box plots
    summarize the trial-level distributions, and black diamonds indicate the
    mean.
    (b) Estimated source-count distribution for the no-\(k\) variant.
    The true number of active sources is \(k=2\).
    }
    \label{fig:cardinality_ablation}
\end{figure}

Figure~\ref{fig:cardinality_ablation} compares the localization performance
of the exact-\(k\), soft-\(k\), and no-\(k\) formulations under the
\(k=2\) condition.

As shown in Figure~\ref{fig:cardinality_ablation}a, the exact-\(k\) and
soft-\(k\) formulations produced similar localization error distributions.
Exact-\(k\) achieved a mean localization error of \(8.92\,\mathrm{mm}\)
and a median error of \(2.08\,\mathrm{mm}\), with an interquartile range
of \([0.00,6.89]\,\mathrm{mm}\). Soft-\(k\) achieved a mean error of
\(9.41\,\mathrm{mm}\) and a median error of \(2.50\,\mathrm{mm}\), with
an interquartile range of \([0.00,6.14]\,\mathrm{mm}\). Both formulations
selected exactly two sources in all 100 trials, yielding a cardinality
accuracy of \(100\%\). These results show that, under the tested condition,
the quadratic soft-cardinality penalty closely reproduced the localization
performance of the exact-\(k\) reference without imposing a hard equality
constraint.

The no-\(k\) formulation retained a low median error but showed a broader
trial-level distribution. Its mean localization error was
\(14.68\,\mathrm{mm}\), with a median of \(2.79\,\mathrm{mm}\) and an
interquartile range of \([0.00,25.00]\,\mathrm{mm}\). The difference in
mean error was therefore driven primarily by a subset of trials with larger
localization errors, including trials in which the selected source count
differed from the true value.

Figure~\ref{fig:cardinality_ablation}b further shows the source-count
variability of the no-\(k\) formulation. It selected the correct source count,
\(\hat{k}=2\), in \(73\%\) of trials. Under-selection to \(\hat{k}=1\)
occurred in \(16\%\) of trials, whereas over-selection occurred with
\(\hat{k}=3\) in \(8\%\) and \(\hat{k}=4\) in \(3\%\) of trials.
The mean estimated source count was \(1.98\), which was close to the true
value \(k=2\), despite a trial-level cardinality accuracy of only \(73\%\).

Overall, the ablation indicates that explicit use of target source-count
information improved the consistency of source-count selection. The primary
soft-\(k\) formulation closely matched the exact-\(k\) reference while
retaining the unconstrained QUBO form. In contrast, removing the target
source-count term increased trial-to-trial variability in both selected
cardinality and localization error.

\FloatBarrier

%% file: tables/table01_main_benchmark_summary_full.tex
\begin{table}[H]
\centering
\caption{
Quantitative summary of the main benchmark under the default simulation
setting. Localization error includes a symmetric source-count mismatch
penalty of \(D_{\mathrm{card}}=50\,\mathrm{mm}\) per mismatched source.
Hit rates are normalized by the number of ground-truth sources.
}
\label{tab:main_benchmark}
\resizebox{\linewidth}{!}{%
\begin{tabular}{lccccc}
\toprule
Method & N & Mean $\pm$ SD (mm) & Median [IQR] (mm) & Hit@10 mm (\%) & Hit@20 mm (\%) \\
\midrule
QUBO & 100 & 8.45 $\pm$ 14.37 & 2.76 [0.00, 7.37] & 84.5 & 90.0 \\
MNE & 100 & 53.95 $\pm$ 25.46 & 53.51 [40.26, 70.00] & 36.5 & 41.0 \\
dSPM & 100 & 26.94 $\pm$ 16.45 & 21.01 [13.15, 40.50] & 22.5 & 68.0 \\
MxNE & 100 & 22.35 $\pm$ 16.99 & 25.00 [6.06, 28.37] & 64.5 & 68.0 \\
RAP-MUSIC & 100 & 25.35 $\pm$ 30.60 & 4.51 [0.00, 48.61] & 60.5 & 67.0 \\
LCMV & 100 & 35.65 $\pm$ 16.51 & 33.47 [22.74, 46.71] & 14.0 & 37.0 \\
\bottomrule
\end{tabular}%
}
\end{table}

%% file: sections/06_discussion.tex
\section{Discussion}
\label{sec:discussion}

This study formulated sparse M/EEG source localization as a binary
source-support selection problem over cortical source groups and constructed
a QUBO formulation with residual-aware candidate screening for joint support
selection. In the controlled synthetic MEG experiments, the proposed method
achieved lower localization error than the evaluated baseline methods under
the default two-source condition. QUBO also maintained favorable performance
across the tested sensor-noise and source-count conditions, although RAP-MUSIC
became competitive in some low-noise and higher-source-count settings. These
results support binary support selection as a useful formulation for sparse
electromagnetic source localization under the conditions considered in this
study.

The differences among methods are consistent with their underlying
source-estimation strategies. MNE estimates a spatially distributed
minimum-norm solution and can therefore produce relatively extended source
patterns. Such estimates are useful for representing broader cortical
activity, but they are not specifically designed for the focal
source-location metric used here. MxNE instead promotes sparse source
activity, while RAP-MUSIC estimates a small set of source locations through
sequential subspace-based fitting. The proposed QUBO formulation differs in
that it performs joint binary selection over a screened set of candidate
source-location groups using candidate-specific sensor-space templates and
their pairwise interactions. The observed performance should therefore be
interpreted in the context of a benchmark that emphasizes recovery of a
small number of focal active locations. For methods producing distributed
source maps, the reported performance also depends on the postprocessing
used to convert those maps into discrete source locations, including the
percentile threshold and spatial linkage distance. Future work should
therefore evaluate sensitivity to these choices and compare methods under a
common parameter-selection protocol.

The candidate-coverage analysis shows that candidate screening is an
important component of the proposed pipeline. Because QUBO optimization is
performed only over the screened candidate set, the quality of this set
directly constrains the final localization result. Trials with complete exact
candidate coverage showed substantially lower localization error than those
with partial coverage. Because this analysis was post hoc, however, it does
not fully separate errors attributable to candidate screening from those
arising during binary optimization. Nevertheless, the observed association
is consistent with candidate omission contributing to some high-error
trials, since a true source excluded from the candidate set cannot be
selected at its exact location. This result identifies candidate generation
and residual-aware screening as important targets for further improvement.
In particular, improving candidate coverage while keeping the candidate set
sufficiently compact could improve localization accuracy without
unnecessarily increasing the size of the subsequent binary optimization
problem.

The cardinality ablation further clarifies the role of target source-count
information. Under the tested \(k=2\) condition, the soft-\(k\) formulation
closely matched the exact-\(k\) reference while retaining an unconstrained
binary optimization form. In contrast, removing the target source-count term
increased trial-to-trial variability in both estimated cardinality and
localization error. These results indicate that reliable prior information
about the expected number of active sources can improve the consistency of
support selection. The soft-cardinality formulation therefore provides a
practical way to incorporate such information without imposing a hard
equality constraint, while still allowing the selected cardinality to be
determined by the QUBO solution. However, the no-\(k\) variant removed
target-count information only from the final QUBO objective, while upstream
template construction still used \(k\). The ablation should therefore be
interpreted as an evaluation of final-stage cardinality control rather than
as a fully \(k\)-independent localization pipeline. Moreover, the reported
localization error includes the symmetric cardinality-mismatch penalty with
\(D_{\mathrm{card}}=50\,\mathrm{mm}\). Comparisons involving methods or
variants with different estimated source counts are therefore conditional
on this penalty and should be interpreted together with the exact-count,
under-selection, and over-selection rates.

The present study was evaluated on controlled synthetic MEG data, and further
validation on real M/EEG measurements and more complex source configurations
will be needed. The QUBO objective also uses fixed candidate-specific
sensor-space templates constructed before binary optimization and should
therefore be viewed as a reduced surrogate for joint continuous coefficient
estimation and support selection rather than an exact reformulation of that
problem. At the same time, the QUBO representation provides a natural
framework for studying larger combinatorial source-selection problems. A
candidate set of size \(\lvert\mathcal{C}\rvert\) produces
\(\lvert\mathcal{C}\rvert\) binary variables and up to
\(O(\lvert\mathcal{C}\rvert^2)\) pairwise QUBO terms, while the number of
possible \(k\)-source supports grows combinatorially with both
\(\lvert\mathcal{C}\rvert\) and \(k\). These scaling considerations motivate
future evaluation of alternative QUBO solvers, including quantum annealing
and hybrid quantum--classical approaches, for larger and more complex
source-localization problems. The present study used classical simulated
annealing and does not make any claim of quantum advantage. Future work can
therefore focus on improving candidate screening, extending validation to
realistic and real-data conditions, evaluating sensitivity to baseline
postprocessing choices, and assessing the scalability of the formulation
across different optimization backends.

%% file: sections/07_conclusion.tex
\section{Conclusion}
\label{sec:conclusion}

This study formulated sparse M/EEG source localization as a binary
source-support selection problem and evaluated the resulting QUBO formulation
with residual-aware candidate screening using controlled synthetic MEG
benchmarks. The proposed approach combines residual-aware candidate screening,
candidate-specific sensor-space templates, pairwise interactions between
candidate templates, and soft source-cardinality control within a binary
quadratic optimization framework. Under the default two-source condition,
the proposed method achieved lower reported localization error than the
evaluated source-localization baselines. It also maintained favorable
localization performance across the tested sensor-noise and source-count
conditions, although RAP-MUSIC became competitive in some low-noise and
higher-source-count settings. Because the reported localization error includes
a symmetric source-count mismatch penalty with
\(D_{\mathrm{card}}=50\,\mathrm{mm}\), comparisons involving different
estimated source counts are conditional on this penalty and should be
interpreted together with the corresponding source-count selection rates.

The candidate-coverage analysis showed that trials with complete exact
candidate coverage had substantially lower localization error than trials with
partial coverage. Because QUBO optimization is restricted to the screened
candidate set, this association highlights candidate quality as an important
factor in the proposed pipeline, although the post hoc analysis does not fully
separate screening errors from those arising during binary optimization. The
cardinality ablation further showed that, under the tested \(k=2\) condition,
the primary soft-\(k\) formulation closely matched the exact-\(k\) reference
while retaining an unconstrained binary optimization form. In contrast,
removing the target source-count term from the final QUBO objective increased
trial-to-trial variability in both estimated cardinality and localization
error. Because upstream template construction still used \(k\), this ablation
evaluates final-stage cardinality control rather than a fully
\(k\)-independent localization pipeline. Taken together, these results
identify candidate screening and the appropriate use of target source-count
information as important components of the proposed source-selection
framework under the tested conditions.

Future work should extend the evaluation to real M/EEG data and more complex
multi-source configurations, while further improving candidate generation and
screening. Sensitivity to the postprocessing used to convert distributed
baseline estimates into discrete source locations should also be investigated
under a common parameter-selection protocol. The computational scalability of
the QUBO formulation warrants further evaluation as both the candidate-set
size and the number of active sources increase. A candidate set of size
\(\lvert\mathcal{C}\rvert\) produces
\(\lvert\mathcal{C}\rvert\) binary variables and up to
\(O(\lvert\mathcal{C}\rvert^2)\) pairwise QUBO terms, while the number of
possible \(k\)-source supports grows combinatorially with both
\(\lvert\mathcal{C}\rvert\) and \(k\). These scaling considerations motivate
future evaluation of alternative classical heuristics, hybrid
quantum--classical methods, and quantum annealing for larger and more complex
source-localization problems. 